\documentclass[twocolumn,showpacs,preprintnumbers,amsmath,amssymb,superscriptaddress]{revtex4}

\usepackage{graphicx}
\usepackage{amssymb}
\newcommand{\di}{\displaystyle}

\begin{document}
\title{Magnetic Frustration in a Mn Honeycomb Lattice Induced 
by Mn-O-O-Mn Pathways}

\author{H.~Wadati}
\email{wadati@ap.t.u-tokyo.ac.jp}
\homepage{http://www.geocities.jp/qxbqd097/index2.htm}
\affiliation{Department of Applied Physics and Quantum-Phase Electronics 
Center (QPEC), University of Tokyo, Hongo, Tokyo 113-8656, Japan} 
\affiliation{Department of Physics and Astronomy, 
University of British Columbia, 
Vancouver, British Columbia V6T 1Z1, Canada}

\author{K.~Kato}
\affiliation{Department of Physics and Department of Complexity Science and Engineering, 
University of Tokyo, Kashiwa, Chiba 277-8561, Japan}

\author{Y.~Wakisaka}
\affiliation{Department of Physics and Department of Complexity Science and Engineering, 
University of Tokyo, Kashiwa, Chiba 277-8561, Japan}

\author{T.~Sudayama}
\affiliation{Department of Physics and Department of Complexity Science and Engineering, 
University of Tokyo, Kashiwa, Chiba 277-8561, Japan}

\author{D. G. Hawthorn}
\affiliation{Department of Physics and Astronomy, 
University of Waterloo, 
Waterloo, Ontario N2L 3G1, Canada}

\author{T.~Z.~Regier}
\affiliation{Canadian Light Source, 
University of Saskatchewan, 
Saskatoon, Saskatchewan S7N 0X4, Canada}

\author{N.~Onishi}
\affiliation{Institute for Chemical Research, 
Kyoto University, Uji, Kyoto 611-0011, Japan}

\author{M.~Azuma}
\altaffiliation[Present address: ]
{Materials and Structures 
Laboratory Tokyo Institute of Technology, 
4259 Nagatsuta, Yokohama 226-8503, Japan}
\affiliation{Institute for Chemical Research, 
Kyoto University, Uji, Kyoto 611-0011, Japan}

\author{Y.~Shimakawa}
\affiliation{Institute for Chemical Research, 
Kyoto University, Uji, Kyoto 611-0011, Japan}

\author{T.~Mizokawa}
\affiliation{Department of Physics and Department of Complexity Science and Engineering, 
University of Tokyo, Kashiwa, Chiba 277-8561, Japan}

\author{A.~Tanaka}
\affiliation{Department of Quantum Matters, ADSM, 
Hiroshima University, Hiroshima 739-8530, Japan}

\author{G.~A.~Sawatzky}
\affiliation{Department of Physics and Astronomy, 
University of British Columbia, 
Vancouver, British Columbia V6T 1Z1, Canada}


\date{\today}
\begin{abstract}
We investigated the electronic structure of 
layered Mn oxide Bi$_3$Mn$_4$O$_{12}$(NO$_3$) 
with a Mn honeycomb lattice by x-ray absorption 
spectroscopy. The valence of Mn was determined to be 
$4+$ with a small charge-transfer energy. 
We estimated the values of 
superexchange interactions up to the fourth nearest 
neighbors ($J_1$, $J_2$, $J_3$, and $J_4$) 
by unrestricted Hartree-Fock calculations and 
a perturbation method. 
We found that the absolute values of $J_1$ through $J_4$ 
are similar with 
positive (antiferromagnetic) $J_1$ and $J_4$, and 
negative (ferromagnetic) $J_2$ and $J_3$, 
due to Mn-O-O-Mn pathways activated by the smallness 
of charge-transfer energy. The negative $J_3$ provides 
magnetic frustration in the honeycomb lattice to prevent long-range ordering. 
\end{abstract}
\maketitle
Since the resonating valence bond state in geometrically 
frustrated magnets has been proposed by Anderson \cite{Anderson1}, 
spin-disordered ground states in Mott insulators 
on frustrated lattices have been attracting great interest 
in condensed-matter physics.
The exchange interaction $J$ in a Mott insulator is roughly 
given by $-2t^2/E_g$, where $t$ is the transfer integral between 
the two localized orbitals and $E_g$ is the excitation energy 
across the Mott gap. 
In Mott insulators on frustrated lattices, 
spin-disordered systems including organic and inorganic materials 
\cite{Shimizu, Nakatsuji, Yamashita, Helton, YOkamoto} 
all have relatively small $E_g$, suggesting that the smallness of $E_g$ 
or the closeness to the Mott transition would be important 
to realize the spin-disordered ground states. 

Various insulating transition-metal oxides are known 
as Mott insulators and can be classified into 
(i) the Mott-Hubbard type insulators 
where the Mott gap $E_g$ is mainly determined 
by the Coulomb interaction $U$ between the transition-metal 
$d$ electrons and 
(ii) the charge-transfer type insulators where $E_g$ is determined 
by the charge-transfer energy $\Delta$ from the oxygen $p$ 
state to the transition-metal $d$ state \cite{ZSA}. 
Therefore, the smallness of $E_g$ can be obtained 
in transition-metal oxides with small $U$ or small $\Delta$. 
In the small $U$ case, theoretical studies on triangular-lattice 
Hubbard models proved that a spin-disordered phase is realized 
near the Mott transition 
\cite{MoritaImada, Motrunich, Senthil, LeeLee}, 
which could be related to the higher order exchange terms. 
As for the small $\Delta$ case, in addition to the higher order terms, 
the exchange pathways through the oxygen $p$ state may give 
unexpectedly long ranged exchange terms and may affect 
the spin disordering.

\begin{figure}
\begin{center}
\includegraphics[width=6cm]{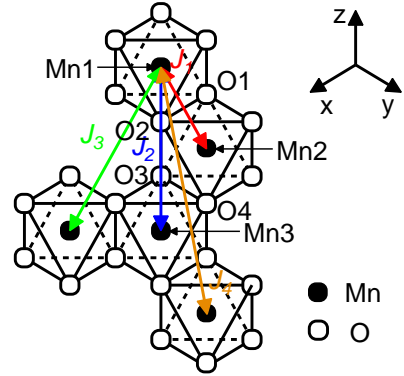}
\caption{(Color online): Network of MnO$_6$ octahedrons in the 
honeycomb lattice with the definitions of 
interactions in 
the nearest-neighbor $J_1$, 
the second nearest-neighbor $J_2$, 
the third nearest-neighbor $J_3$, and 
the fourth nearest neighbor $J_4$.}
\label{fig3}
\end{center}
\end{figure}

Very recently, a spin-disordered ground state is 
reported in a layered Mn oxide Bi$_3$Mn$_4$O$_{12}$(NO$_3$) 
with a Mn honeycomb lattice in which the exchange 
interaction between second neighbor Mn sites introduces 
a kind of frustration in the honeycomb lattice \cite{azumahoney}. 
In this material, there is a network of MnO$_6$ 
octahedrons and Fig.~\ref{fig3} shows the definitions of 
interactions in the nearest-neighbor $J_1$, 
the second nearest-neighbor $J_2$, 
the third nearest-neighbor $J_3$, and 
the fourth nearest neighbor $J_4$. 
The exchange interaction $J_2$ between second neighbor sites 
could be derived from the Mn-O-O-Mn exchange pathways which becomes 
important in small $\Delta$ systems. 
The values of these interactions were studied both 
theoretically \cite{motome} and 
experimentally by inelastic neutron scattering \cite{nt}. 
In these studies, only $J_1$ and $J_2$ are considered 
by assuming that the values of $J_3$ and $J_4$ are small 
enough to be neglected. 
However, since the exchange interaction $J_3$ 
between the third neighbor sites is also derived 
from  the Mn-O-O-Mn exchange pathways, 
it is not a trivial question whether $J_3$ 
is negligible compared to $J_2$ or not. 
In this context, it is very interesting and important to study 
the electronic structure of the Mn oxide, especially 
the values of interactions $J_1$ - $J_4$, 
using spectroscopic methods 
and to reveal the origin of the spin-disordered state 
from the electronic structural viewpoint. 
In this paper, 
we investigated the electronic structure of this material 
by x-ray absorption spectroscopy (XAS), and also 
estimated the values of magnetic interactions. 
Unrestricted Hartree-Fock calculations and a perturbation method 
revealed that the nearest neighbor and fourth nearest neighbor 
$J_1$ and $J_4$ are positive (antiferromagnetic), and 
the next nearest neighbor and third nearest neighbor 
$J_2$ and $J_3$ are negative (ferromagnetic). 
In the present analysis, the ferromagnetic $J_2$ 
does not introduce magnetic frustration in the honeycomb lattice. 
We conclude that the ferromagnetic $J_3$ is 
the origin of magnetic frustration and 
the absence of long-range ordering in this material. 

The synthesis of Bi$_3$Mn$_4$O$_{12}$(NO$_3$) 
polycrystalline powder is described in Ref.~\cite{azumahoney}. 
X-ray absorption experiments were performed 
at 11ID-1 (SGM) of the Canadian Light Source. 
The spectra were measured in the total-electron-yield 
(TEY) mode. The total energy resolution 
was set to 100 meV. All the spectra were measured 
at room temperature. 
The obtained spectrum is analyzed by standard 
cluster-model calculations  \cite{atanaka2} 
to obtain electronic parameters. 
The parameters in this model are 
$3d$ - $3d$ and $3d$ - $2p$ 
Coulomb interactions ($U_{dd}$ and $U_{dc}$, 
respectively), charge-transfer energy from O $2p$ 
to Mn $3d$ states $\Delta$, hopping integrals 
between Mn $3d$ and O $2p$ molecular 
states [$V(t_{2g})$ and $V(e_g)$], 
and crystal field parameter 
$10Dq$. The superexchange interactions are evaluated 
using unrestricted Hartree-Fock calculation 
with a multi-band $d-p$ Hamiltonian 
with Mn 3$d$ and O 2$p$ states \cite{Mizokawa1996}. 
The Hamiltonian is given by 
\begin{eqnarray*}
H = H_p + H_d + H_{pd},
\end{eqnarray*}

\begin{eqnarray*}
H_p = \sum_{k,l,\sigma}
\epsilon^p_{k} p^+_{k,l\sigma}p_{k,l\sigma}
+ \sum_{k,l>l',\sigma}
V^{pp}_{k,ll'} p^+_{k,l\sigma}p_{k,l'\sigma}
+ H.c.,
\end{eqnarray*}

\begin{eqnarray*}
H_d & = &\sum_{k,m\sigma}\epsilon_d d^+_{k,m\sigma}d_{k,m\sigma}\\
& + &
 \sum_{k,m>m',\sigma}V^{dd}_{k,mm'}d^+_{k,m\sigma}d_{k,m'\sigma}+H.c.\\
& + &u\sum_{i,m}d^+_{i,m\uparrow}d_{i,m\uparrow}d^+_{i,m\downarrow}d_{i,m\downarrow}\\
& + &u'\sum_{i,m\neq m'}d^+_{i,m\uparrow}d_{i,m\uparrow}
d^+_{i,m'\downarrow}d_{i,m'\downarrow}\\
\end{eqnarray*}
\begin{eqnarray*}
& + &(u'-j')\sum_{i,m>m',\sigma}d^+_{i,m\sigma}d_{i,m\sigma}
d^+_{i,m'\sigma}d_{i,m'\sigma}\\
& + &j'\sum_{i,m\neq m'}d^+_{i,m\uparrow}d_{i,m'\uparrow}
d^+_{i,m\downarrow}d_{i,m'\downarrow}\\
& + &j\sum_{i,m\neq m'}d^+_{i,m\uparrow}d_{i,m'\uparrow}
d^+_{i,m'\downarrow}d_{i,m\downarrow},
\end{eqnarray*}

\begin{eqnarray*}
H_{pd} = \sum_{k,m,l,\sigma} V^{pd}_{k,lm}
d^+_{k,m\sigma}p_{k,l\sigma} + H.c.
\end{eqnarray*}

\noindent Here, $d^+_{i,m\sigma}$ are creation operators for the Mn 3$d$ 
electrons at site $i$. $d^+_{k,m\sigma}$ and $p^+_{k,l\sigma}$ are creation 
operators for Bloch electrons with momentum $k$ which are 
constructed from the $m$-th component of the Mn 3$d$ orbitals 
and from the $l$-th component of the O 2$p$ orbitals, respectively.
The intra-atomic Coulomb interaction between the Mn 3$d$ electrons
is expressed using Kanamori parameters, $u$, $u'$, $j$ and $j'$
satisfying the relations $u = u'+j+j'$ and $j=j'$. 
The transfer integrals between the Mn 3$d$ and O 2$p$ orbitals 
$V^{pd}_{k,lm}$ are given in terms of Slater-Koster parameters 
$(pd\sigma)$ and $(pd\pi)$.
The parameters determined by the cluster-model calculation 
are used as input of the unrestricted Hartree-Fock analysis. 

\begin{figure}
\begin{center}
\includegraphics[width=6cm]{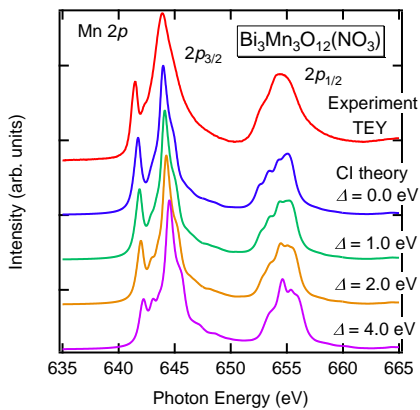}
\caption{(Color online): 
Mn $2p$ XAS spectra of Bi$_3$Mn$_4$O$_{12}$(NO$_3$) 
and comparison with the CI theory.}
\label{fig1}
\end{center}
\end{figure}

Figure \ref{fig1} shows the Mn $2p$ 
XAS spectrum of Bi$_3$Mn$_4$O$_{12}$(NO$_3$). 
There are two structures, 
Mn 2$p_{3/2}$ $\rightarrow$ $3d$ absorption at 640 - 650 eV and 
Mn $2p_{1/2}$ $\rightarrow$ $3d$ absorption at 650 - 660 eV. 
The experimental spectrum has a sharp peak at $\sim 641.5$ eV 
characteristic of Mn$^{4+}$, 
concluded by comparing with the reference data of Mn$^{2+}$ (MnO), 
Mn$^{3+}$ (LaMnO$_3$), and Mn$^{4+}$ 
(EuCo$_{0.5}$Mn$_{0.5}$O$_3$ and SrMnO$_3$) 
from Ref.~\cite{MnCo}. 
This indicates that the valence of Mn is $4+$ in 
Bi$_3$Mn$_4$O$_{12}$(NO$_3$), consistent 
with the valance state of 
Bi$^{3+}_3$Mn$^{4+}_4$O$^{2-}_{12}$(NO$_3$)$^-$ 
obtained in Ref.~\cite{azumahoney}. 
We performed configuration-interaction (CI) cluster-model 
calculations \cite{atanaka2} to obtain electronic 
parameters. Here we fixed the following values 
$U_{dd} = 6.0$ eV, $U_{dc}$ = 7.5 eV, 
$V(e_g) = 3.0$ eV, and $10 Dq =1.3$ eV 
and changed the value of $\Delta$ 
from 0.0 eV to 4.0 eV, as shown in 
Fig.~\ref{fig1} (a). 
The calculated spectra do not depend 
on the value of $\Delta$ so much, but 
the small $\Delta$ values 
of 1.0 $\pm$ 1.0 eV reproduce 
the experiment most successfully 
from a peak at $\sim 641.5$ eV 
and Mn $2p_{1/2}$ structures. 

\begin{figure}
\begin{center}
\includegraphics[width=7cm]{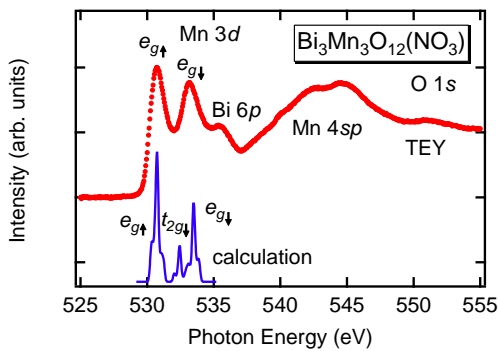}
\caption{(Color online): 
O $1s$ XAS spectra of Bi$_3$Mn$_4$O$_{12}$(NO$_3$) together 
with the calculated O 1$s$ partial density of states 
by unrestricted Hartree-Fock calculations.}
\label{fig2}
\end{center}
\end{figure}

Figure \ref{fig2} shows the O $1s$ XAS spectrum 
of Bi$_3$Mn$_4$O$_{12}$(NO$_3$) and 
the calculated O 1$s$ partial density of states 
by unrestricted Hartree-Fock calculations. 
There are two structures in the 
O $2p$ - Mn $3d$ hybridized states, 
assigned as 
majority-spin $e_g$ states ($e_{g\uparrow}$) 
and 
minority-spin $e_{g}$ states ($e_{g\downarrow}$), 
and another structure in the 
O $2p$ - Bi $6p$ states as shown in Fig.~\ref{fig2}. 
There is a higher intensity at $e_{g\downarrow}$ 
states in the experiment than in the calculation, 
indicating that there is a substantial Bi $6p$ 
contribution also in this structure. 
These assignments are consistent with the 
Mn$^{4+}$ ($d^3$) state, where majority-spin 
$t_{2g}$ states are occupied by electrons. 

In order to obtain the values of magnetic interactions, 
we have performed 
unrestricted Hartree-Fock analysis on the multiband $d-p$ 
Hamiltonian with the Mn 3$d$ and O 2$p$ orbitals. 
$\Delta$, $U$, and ($pd\sigma$) were set to be 1.0, 6.0, 
and $-1.8$ eV, respectively, on the basis of the cluster 
model analysis. The ratio ($pd\sigma$)/($pd\pi$) is $-$2.16. 
Remaining transfer integrals expressed by ($pp\sigma$), 
($pp\pi$), ($dd\sigma$), and ($dd\pi$) are fixed 
at $-$0.6, 0.15, $-$0.3, and 0.15 eV, respectively, for the 
honeycomb lattice with the regular MnO$_6$ 
octahedron. Here $\Delta$ denotes the charge-transfer energy 
for Mn$^{4+}$ specifically 
${\Delta} = {\epsilon}_d - {\epsilon}_p + 3U$. 

With this parameter set, the lowest energy state 
is found to be the conventional antiferromagnetic state 
where the first neighboring sites are antiferromagnetically coupled. 
In the antiferromagnetic ground state, 
the second, third, and forth neighboring sits are ferromagnetically, 
antiferromagnetically, and antiferromagnetically coupled, respectively. 
We have calculated energies of ferromagnetic state as well as 
modified antiferromagnetic states which are obtained by exchanging spins 
of some neighboring sites in the ground state. 
As expected, the ferromagnetic state is very much higher in energy 
than the antiferromagnetic ground state. However, some of the modified 
antiferromagnetic states were found to have energies very close to 
that of the ground state. By mapping the Hartree-Fock energies to 
the Heisenberg model with  $J_1$, $J_2$, $J_3$, and $J_4$, 
the obtained values are:
\begin{equation}
\begin{array}{rcl}
J_1 & =  & 9.15 \ \mbox{meV}\\
J_2 & =  & -5.32  \ \mbox{meV}\\
J_3 & = & -4.80  \ \mbox{meV}\\
J_4 & = & 5.77 \ \mbox{meV}
\end{array}
\end{equation}

While the antiferromagnetic $J_1$ and ferromagnetic $J_2$, and 
antiferromagnetic $J_4$ are consistent with the antiferromagnetic ground 
state, the ferromagnetic $J_3$ can introduce frustration effect on it. 
Since the magnitude of the ferromagnetic $J_3$ is comparable to those 
of the antiferromagnetic $J_1$ and ferromagnetic $J_2$, the ferromagnetic 
$J_3$ by the Mn-O-O-Mn superexchange pathways is responsible for 
the absence of long-range ordering in the present honeycomb system. 

Let us also examine the sign of $J_3$ by considering 
superexchange pathways in Bi$_{3}$Mn$_{4}$O$_{12}$(NO$_3$) 
using a perturbation method 
based on the electronic structure parameters 
obtained from the analysis of the Mn $2p$ XAS spectrum. 
Here, we will use Slater-Koster parameters, 
$(pp\sigma)$, $(pd\sigma)$, $(dd\sigma)$, 
and so on \cite{JC}. 

$J_3$ is given as  
\begin{equation}
\begin{array}{rcl} 
J_3 & = & 
\left(
\di\frac{(pd\pi)^4}{[\Delta-(pp\sigma)/2+(pp\pi)/2]^2}+
\right.\\
& & 
\left.\di\frac{(pd\pi)^4}{[\Delta+(pp\sigma)/2-(pp\pi)/2]^2}
\right)\left(\di\frac{1}{\Delta+u_p}+\frac{1}{u}\right)
\\
 & & 
+\di\frac{2(pd\pi)^4}
{[\Delta-(pp\sigma)/2+(pp\pi)/2][\Delta+(pp\sigma)/2-(pp\pi)/2]} \times\\
 & & 
\left(\di\frac{1}{\Delta+u_p}+\frac{1}{u}-\frac{1}{\Delta+u_p-j_p}\right)\\
 & & 
-\left(
\di\frac{(pd\sigma)^2(pd\pi)^2}{[\Delta-(pp\sigma)/2-(pp\pi)/2]^2}+\right. \\
 & & \left.
\di\frac{(pd\sigma)^2(pd\pi)^2}{[\Delta+(pp\sigma)/2+(pp\pi)/2]^2}
\right) \times \\
& & 
\left(
\di\frac{1}{\Delta+u_p}+\frac{1}{u-3j}-\frac{1}{u-2j}
\right)\\
 & & 
-\di\frac{2(pd\sigma)^2(pd\pi)^2}
{[\Delta-(pp\sigma)/2-(pp\pi)/2][\Delta+(pp\sigma)/2+(pp\pi)/2]}
\times \\ 
 & & 
\left(
\di\frac{1}{\Delta+u_p}+\frac{1}{u-3j}-\frac{1}{\Delta+u_p-j_p}
-\frac{1}{u-2j}
\right)\\\\
 & \sim & 
\di\frac{2(pd\pi)^4}{\Delta^2}
\left(1-\di\frac{[(pp\sigma)/2-(pp\pi)/2]^2}{\Delta^2}\right)
\frac{1}{\Delta+u_p}\\
 & & 
+\di\frac{4(pd\pi)^4[(pp\sigma)/2-(pp\pi)/2]^2}{\Delta^4}
\left(
\di\frac{1}{u}-\frac{j_p}{(\Delta+u_p)^2}
\right)\\
 & & 
-\di\frac{2(pd\sigma)^2(pd\pi)^2}{\Delta^2}
\left(1-\frac{[(pp\sigma)/2+(pp\pi)/2]^2}{\Delta^2}\right)
\frac{1}{\Delta+u_p}\\
& & 
-\di\frac{4(pd\sigma)^2(pd\pi)^2[(pp\sigma)/2+(pp\pi)/2]^2}{\Delta^2} 
\times \\
& & 
\left(\di\frac{j}{u^2}-\frac{j_p}{(\Delta+u_p)^2}\right)
\end{array}
\end{equation}
Here we considered 
the molecular orbitals made from two oxygen sites. 
The first and second terms proportional to 
($pd\pi$)$^4$ are given by 
the pathway from the Mn $t_{2g}$ states 
to the O 2$p$ molecular orbitals 
to the Mn $t_{2g}$ states. 
The third and fourth terms proportional to 
($pd\sigma$)$^2$($pd\pi$)$^2$ are given by the pathway from 
the Mn $e_g$ states to the O 2$p$ molecular orbitals 
to the Mn $t_{2g}$ states. 
$J_3$ is dominated by the negative third term 
and causes ferromagnetic interactions, which 
is consistent with the result of the unrestricted 
Hartree Fock calculations. 

From the neutron measurements, it was found that 
the values of interlayer interactions ($J_c$) 
are also comparable to $J_1$  \cite{nt}. 
This is consistent with our result because 
$J_c$ is also determined by Mn-O-O-Mn pathways. 

We investigated the electronic structure of 
Bi$_3$Mn$_4$O$_{12}$(NO$_3$) by XAS. 
The valence of Mn was determined to be 
$4+$, and from CI theory we found that 
a charge-transfer energy is small in this material. 
Then we estimated the values of $J_{1,2,3,4}$ 
by unrestricted Hartree-Fock calculations and 
a perturbation method. 
We found antiferromagnetic $J_1$ and ferromagnetic $J_2$ and $J_3$, 
leading to the existence of magnetic frustration and 
the absence of long-range ordering, 
as experimentally confirmed. 

The authors would like to thank Y. Motome for 
informative discussions. 
This research was made possible with financial 
support from the Canadian funding organizations 
NSERC, CFI, and CIFAR. H.W. is supported by 
the Japan Society for the Promotion 
of Science (JSPS) through its Funding Program for 
World-Leading Innovative R{\&}D on Science and Technology 
(FIRST Program). 

\bibliography{LVO1tex}
\end{document}